\documentclass[aps,prl,reprint,superscriptaddress,showkeys,showpacs]{revtex4-1}

\usepackage{graphicx} %
\usepackage{amsmath} %
\usepackage{amssymb} %
\usepackage{subfigure} %
\usepackage{float} %
\usepackage{upgreek} %
\usepackage{multirow} %
\usepackage{color} %
\usepackage{lineno} %
\usepackage{hyperref} %
\usepackage{booktabs}
\usepackage{placeins}
\usepackage[normalem]{ulem}
\usepackage[usenames,dvipsnames]{xcolor}
\hypersetup{colorlinks=true, urlcolor=blue, linkcolor=blue, citecolor=red} %
\usepackage{lineno}

\begin{document}


\newcommand{\dm}{\mbox{DM-Ice}}
\newcommand{\dmi}{\mbox{DM-Ice17}}

\newcommand{\iso}[2]{\ensuremath{^{#2}\mathrm{#1}}}

\newcommand{\dma}{\mbox{Det-1}}
\newcommand{\dmb}{\mbox{Det-2}}
\newcommand{\pmta}{\mbox{PMT-1a}}
\newcommand{\pmtb}{\mbox{PMT-1b}}
\newcommand{\pmtc}{\mbox{PMT-2a}}
\newcommand{\pmtd}{\mbox{PMT-2b}}

\newcommand{\modlimit}{[LIMIT]}
\newcommand{\EnRange}{4\,--\,20\,\ensuremath{\mathrm{keV}}}
\newcommand{\EnRangeA}{4\,--\,20\,\ensuremath{\mathrm{keV}}}
\newcommand{\EnRangeB}{6\,--\,20\,\ensuremath{\mathrm{keV}}}
\newcommand{\exposurekgdays}{21,717.5\,kg$\cdot$days}
\newcommand{\exposurekgyrs}{60.8\,kg$\cdot$yr}
\newcommand{\analysisthreshold}{4\,\ensuremath{\mathrm{keV}}}
\newcommand{\analysisthresholdA}{4\,\ensuremath{\mathrm{keV}}}
\newcommand{\analysisthresholdB}{6\,\ensuremath{\mathrm{keV}}}

\newcommand{\startmonth}{June 2011}
\newcommand{\stopmonth}{January 2015}
\newcommand{\startdate}{June 16, 2011}
\newcommand{\stopdate}{January 28, 2015}
\newcommand{\totalhours}{31442} 
\newcommand{\totaldays}{1310.1} 

\newcommand{\black}[1]{{\color{black}{#1}}}
\newcommand{\red}[1]{{\color{Red}{#1}}}
\newcommand{\blue}[1]{{\color{Blue}{#1}}}
\newcommand{\green}[1]{{\color{Green}{#1}}}
\newcommand{\gray}[1]{{\color{gray}{#1}}}
\newcommand{\orange}[1]{{\color{BurntOrange}{#1}}}

\newcommand{\h}{ht}
\newcommand{\fw}{\textwidth}
\newcommand{\tqw}{0.75\textwidth}
\newcommand{\hw}{0.48\textwidth} 
\newcommand{\qw}{0.23\textwidth}
\newcommand{\tw}{0.31\textwidth}

\newcommand{\fig}[1]{Fig.\,\ref{#1}}
\newcommand{\tab}[1]{Tab.\,\ref{#1}}
\newcommand{\sect}[1]{Sec.\,\ref{#1}}
\newcommand{\chap}[1]{Chap.\,\ref{#1}}
\newcommand{\equ}[1]{Eq.\,\ref{#1}}
\newcommand{\cit}[1]{Ref.\,\cite{#1}}

\renewcommand{\deg}{\ensuremath{{^{\circ}}\mathrm{C}}}

\newcommand{\aprox}{\ensuremath{\sim}}
\newcommand{\per}{\,\%}
\newcommand{\kpa}{\,kPa}
\newcommand{\hz}{\,Hz}
\newcommand{\s}{\,sec}
\newcommand{\hr}{\,hours}
\newcommand{\lt}{\ensuremath{\,<\,}}
\newcommand{\gt}{\ensuremath{\,>\,}}
\newcommand{\plm}{\ensuremath{\,\pm}\,}
\newcommand{\kg}{\,kg}
\newcommand{\kgyr}{\,kg\ensuremath{\cdot}yr}
\newcommand{\gs}{\ensuremath{\,g}}
\newcommand{\diam}{\ensuremath{\varnothing\,}}
\newcommand{\usec}{\,\ensuremath{\mathrm{\upmu s}}}
\newcommand{\nsec}{\,\ensuremath{\mathrm{ns}}}
\newcommand{\msec}{\,\ensuremath{\mathrm{ms}}}
\newcommand{\ubk}{\,\ensuremath{\upmu}Bq/kg}
\newcommand{\mbk}{\,mBq/kg}

\newcommand{\kev}{\,\ensuremath{\mathrm{keV}}}
\newcommand{\mev}{\,\ensuremath{\mathrm{MeV}}}
\newcommand{\kevee}{\,\ensuremath{\mathrm{keV}_{\mathrm{ee}}}}
\newcommand{\kevnr}{\,\ensuremath{\mathrm{keV}_{\mathrm{nr}}}}
\newcommand{\dru}{\,\ensuremath{\mathrm{counts/day/keV/kg}}}

\newcommand{\gev}{\,\ensuremath{{\rm GeV}/{\rm c}^2}}
\newcommand{\smod}{\ensuremath{{S_m}}}
\newcommand{\snaught}{\ensuremath{{S_0}}}
\newcommand{\rzero}{\ensuremath{{R_0}}}
\newcommand{\nzero}{\ensuremath{{N_0}}}

\newcommand{\e}[1]{\ensuremath{\times 10^{#1}}}

\newcommand{\kms}{\,\ensuremath{\mathrm{km/s}}}

\title{First search for a dark matter annual modulation signal with NaI(Tl) in the Southern Hemisphere by \dmi}

\author {E.~Barbosa de Souza}
\affiliation {Department of Physics and Wright Laboratory, Yale University, New Haven, Connecticut 06520, USA}

\author {J.~Cherwinka}
\affiliation {Physical Sciences Laboratory, University of Wisconsin-Madison, Stoughton, Wisconsin 53589, USA}

\author {A.~Cole}
\affiliation {Department of Physics and Astronomy, University of Sheffield, Sheffield S10 2TN, United Kingdom}
\affiliation {STFC Boulby Underground Science Facility, Boulby Mine, Cleveland TS13 4UZ, United Kingdom}

\author {A.~C.~Ezeribe}
\affiliation {Department of Physics and Astronomy, University of Sheffield, Sheffield S10 2TN, United Kingdom}

\author {D.~Grant}
\affiliation {Department of Physics, University of Alberta, Edmonton, Alberta T6G 2E1, Canada}

\author {F.~Halzen}
\affiliation {Department of Physics and Wisconsin IceCube Particle Astrophysics Center, University of Wisconsin-Madison, Madison, Wisconsin 53706, USA}

\author {K.~M.~Heeger}
\affiliation {Department of Physics and Wright Laboratory, Yale University, New Haven, Connecticut 06520, USA}

\author {L.~Hsu}
\affiliation {Fermi National Accelerator Laboratory, Batavia, Illinois 60510, USA}

\author {A.~J.~F.~Hubbard}
\altaffiliation [Present address: ]{Department of Physics and Astronomy, Northwestern University, Evanston, Illinois 60208, USA}
\affiliation {Department of Physics and Wright Laboratory, Yale University, New Haven, Connecticut 06520, USA}
\affiliation {Department of Physics and Wisconsin IceCube Particle Astrophysics Center, University of Wisconsin-Madison, Madison, Wisconsin 53706, USA}

\author {J.~H.~Jo}
\affiliation {Department of Physics and Wright Laboratory, Yale University, New Haven, Connecticut 06520, USA}

\author {A.~Karle}
\affiliation {Department of Physics and Wisconsin IceCube Particle Astrophysics Center, University of Wisconsin-Madison, Madison, Wisconsin 53706, USA}

\author {M.~Kauer}
\affiliation {Department of Physics and Wright Laboratory, Yale University, New Haven, Connecticut 06520, USA}
\affiliation {Department of Physics and Wisconsin IceCube Particle Astrophysics Center, University of Wisconsin-Madison, Madison, Wisconsin 53706, USA}

\author {V.~A.~Kudryavtsev}
\affiliation {Department of Physics and Astronomy, University of Sheffield, Sheffield S10 2TN, United Kingdom}

\author {K.~E.~Lim}
\affiliation {Department of Physics and Wright Laboratory, Yale University, New Haven, Connecticut 06520, USA}

\author {C.~Macdonald}
\affiliation {Department of Physics and Astronomy, University of Sheffield, Sheffield S10 2TN, United Kingdom}

\author {R.~H.~Maruyama}
\email[Corresponding author: ]{reina.maruyama@yale.edu}
\affiliation {Department of Physics and Wright Laboratory, Yale University, New Haven, Connecticut 06520, USA}

\author {F.~Mouton}
\affiliation {Department of Physics and Astronomy, University of Sheffield, Sheffield S10 2TN, United Kingdom}

\author {S.~M.~Paling}
\affiliation {STFC Boulby Underground Science Facility, Boulby Mine, Cleveland TS13 4UZ, United Kingdom}

\author {W.~Pettus}
\altaffiliation [Present address: ]{CENPA and Department of Physics, University of Washington, Seattle, Washington 98195, USA}
\affiliation {Department of Physics and Wright Laboratory, Yale University, New Haven, Connecticut 06520, USA}
\affiliation {Department of Physics and Wisconsin IceCube Particle Astrophysics Center, University of Wisconsin-Madison, Madison, Wisconsin 53706, USA}

\author {Z.~P.~Pierpoint}
\email[Corresponding author: ]{zachary.pierpoint@yale.edu}
\affiliation {Department of Physics and Wright Laboratory, Yale University, New Haven, Connecticut 06520, USA}
\affiliation {Department of Physics and Wisconsin IceCube Particle Astrophysics Center, University of Wisconsin-Madison, Madison, Wisconsin 53706, USA}

\author {B.~N.~Reilly}
\altaffiliation {Present address: Department of Physics and Astronomy, University of Wisconsin-Fox Valley, Menasha, Wisconsin 54952, USA}
\affiliation {Department of Physics and Wright Laboratory, Yale University, New Haven, Connecticut 06520, USA}
\affiliation {Department of Physics and Wisconsin IceCube Particle Astrophysics Center, University of Wisconsin-Madison, Madison, Wisconsin 53706, USA}

\author {M.~Robinson}
\affiliation {Department of Physics and Astronomy, University of Sheffield, Sheffield S10 2TN, United Kingdom}

\author {F.~R.~Rogers}
\altaffiliation [Present address: ]{Laboratory for Nuclear Science, Massachusetts Institute of Technology, Cambridge, Massachusetts 02139, USA}
\affiliation {Department of Physics and Wright Laboratory, Yale University, New Haven, Connecticut 06520, USA}

\author {P.~Sandstrom}
\affiliation {Department of Physics and Wisconsin IceCube Particle Astrophysics Center, University of Wisconsin-Madison, Madison, Wisconsin 53706, USA}

\author {A.~Scarff}
\affiliation {Department of Physics and Astronomy, University of Sheffield, Sheffield S10 2TN, United Kingdom}

\author {N.~J.~C.~Spooner}
\affiliation {Department of Physics and Astronomy, University of Sheffield, Sheffield S10 2TN, United Kingdom}

\author {S.~Telfer}
\affiliation {Department of Physics and Astronomy, University of Sheffield, Sheffield S10 2TN, United Kingdom}

\author {L.~Yang}
\affiliation {Department of Physics, University of Illinois at Urbana-Champaign, Urbana, Illinois 61801, USA}
\collaboration{The DM-Ice Collaboration}
\date{\today}

\begin{abstract}
We present the first search for a dark matter annual modulation signal in the Southern Hemisphere conducted with NaI(Tl) detectors, performed by the \dmi\ experiment. Nuclear recoils from dark matter interactions are expected to yield an annually modulated signal independent of location within the Earth's hemispheres. \dmi, the first step in the \dm\ experimental program, consists of 17\kg\ of NaI(Tl) located at the South Pole under 2200\,m.w.e.\ overburden of Antarctic glacial ice. Taken over 3.6 years for a total exposure of \exposurekgyrs, \dmi\ data are consistent with no modulation in the energy range of 4--20\kev, providing the strongest limits on weakly interacting massive particle dark matter from a direct detection experiment located in the Southern Hemisphere. The successful deployment and stable long-term operation of \dmi\ establishes the South Pole ice as a viable location for future dark matter searches and in particular for a high-sensitivity NaI(Tl) dark matter experiment to directly test the DAMA/LIBRA claim of the observation of dark matter.

\keywords{DM-Ice, dark matter, sodium iodide, WIMP, direct detection, annual modulation, South Pole}
\pacs{95.35.+d, 29.40.Mc, 95.55.Vj}

\end{abstract}

\maketitle


Cosmic observations indicate that the majority of the mass in the Universe is not accounted for by the standard model of particle physics~\cite{Ade:2013zuv}. Theoretical models favor the weakly interacting massive particle (WIMP) as an explanation for this dark matter. The WIMP dark matter model implies the existence of at least one additional particle with a mass on the GeV to TeV scale and interaction cross section with normal matter on the nano- to zeptobarn scale~\cite{Steigman:1984ac, Bertone}. The Milky Way is modeled to be surrounded by a halo of dark matter, providing a continuous flux of dark matter through terrestrial detectors~\cite{Scattering1, Scattering2}.

In the standard (WIMP) halo model, the motion of the Solar System through the galaxy produces a net velocity relative to the dark matter halo. As the Earth orbits the Sun, the magnitude of the Earth's velocity parallel to the solar velocity changes. As a result, the apparent velocity of the dark matter wind on Earth modulates with a one year period, leading in turn to an annual modulation in the rate of dark matter interactions~\cite{Scattering2, Freese:1987wu}. 

A suite of experiments attempting to directly detect WIMP dark matter is underway using a variety of detector technologies~\cite{Direct_Detection}. The DAMA/NaI and DAMA/LIBRA experiments have reported observations of an annual modulation signal in thallium-doped sodium iodide [NaI(Tl)] crystals consistent with WIMP dark matter, with the cumulative signature reaching \ensuremath{9.3\sigma}\,C.L.~\cite{DAMANai, DAMA_full}. Under the assumption of the standard dark matter halo model, the constraints for many of the WIMP and WIMP-like models set by other experiments are inconsistent with DAMA's claim for discovery~\cite{Freese, Savage_1, LUX2016, LUXSD, XENON2016, XENON2015ER, CDMS, CRESST2015, PandaX-II, PICO, SIMPLE, Kim:2012rza, XMASS}. Many alternative explanations coupling the modulation to seasonal effects have been proposed~\cite{Ralston, Blum, Davis, Nygren}, but to date, none have been able to fully explain the DAMA signal~\cite{No_role2, FM, Klinger, DMIceMuons}. 

The \dm\ experiment aims to resolve this tension by performing an unambiguous test of the DAMA signal using the same target medium~\cite{DMIce2012}. The expected dark matter modulation has the same phase everywhere on Earth, whereas any modulation resulting from seasonal effects reverses its phase between the Northern and Southern Hemispheres. In order to decouple the dark matter phase from seasonal variations, the \dm\ Collaboration is working towards a Southern Hemisphere deployment of a large NaI(Tl) array. The \dm's location provides a unique complement to other proposed NaI-based experiments to test DAMA's assertion of the detection of dark matter~\cite{ANAIS2016, COSINUS, KIMS_NaI, SABRE}.

\dmi, the first stage of the \dm\ experimental program, was deployed at the geographic South Pole in December 2010~\cite{DMIce2014}. Two 8.47\kg\ crystals originally used in the NAIAD dark matter experiment at the Boulby Underground Laboratory~\cite{NAIAD2} were placed on separate strings at the bottom of the IceCube detector array~\cite{IceCube}, with 2457\,m (2200\,m.w.e.) overburden from the Antarctic ice. \dmi\ constitutes the first search for annual modulation dark matter signal with NaI(Tl) detectors in the Southern Hemisphere. The only previous direct detection dark matter search in the Southern Hemisphere, which includes a modulation analysis, was performed at Sierra Grande in germanium detectors~\cite{SierraGrande}. The primary goal of \dmi\ is to verify the feasibility of performing a direct detection dark matter search in the glacial ice at the South Pole, as the small detector mass and high internal backgrounds preclude a sensitive test of the DAMA signal. The successful deployment and stable data taking with \dmi\ establish the South Pole as a competitive site for future dark matter searches.

The analysis presented here covers the 3.6 year primary physics run of \dmi\ from \startdate\ to \stopdate. This corresponds to a total exposure of \exposurekgyrs~with \gt99\% uptime. Each \dmi\ crystal is optically coupled to two photomultiplier tubes (PMTs) which detect the scintillation photons. These two assemblies are referred to as \dma\ and \dmb. Coincidence between paired PMTs, imposed by a trigger condition, provides an online dark noise reduction. For the dark matter analysis, data are recorded into \aprox600\nsec\ waveforms. Detector details, including the trigger conditions, DAQ configuration, and energy calibration, are described in \cite{DMIce2014}.

Energy information is obtained by integrating the charge over the entire readout. Calibration is performed \textit{in situ} by performing a linear regression between the fitted internal gamma peak mean charge and the corresponding simulated peak energies~\cite{DMIce2014}. In this paper, all energies are presented in electron-equivalent energies and referred to as keV for simplicity. The crystal response to gamma rays is observed to be linear in two distinct regions (0\,--\,100\kev\ and 500\,--\,3000\kev), consistent with previous measurements~\cite{Engelkemeir, Khodyuk}. We observe that the total collected charge associated with a given decay is decreasing over time by \aprox0.5\% per year; the cause of this is under further investigation. The calibration includes a time dependence to account for this apparent gain shift. 

Operating conditions are monitored by sensors on the DAQ mainboards, and their stability is described in \cite{DMIce2014, Pettus}. In addition to the monitoring data and normal physics runs, periodic runs without the PMT coincidence or trigger threshold requirements are taken for detector characterization, particularly characterization of the single photoelectron (SPE) response. Apart from the aforementioned gain shift and a 0.3\deg\ cooling of the detector as the drill ice column thermalizes with the surrounding ice~\cite{Pettus}, all detector behaviors, such as trigger threshold and DAQ deadtime, exhibit no time dependence.

\begin{figure}[t]
	\begin{center}
		\includegraphics[width=0.50\textwidth]{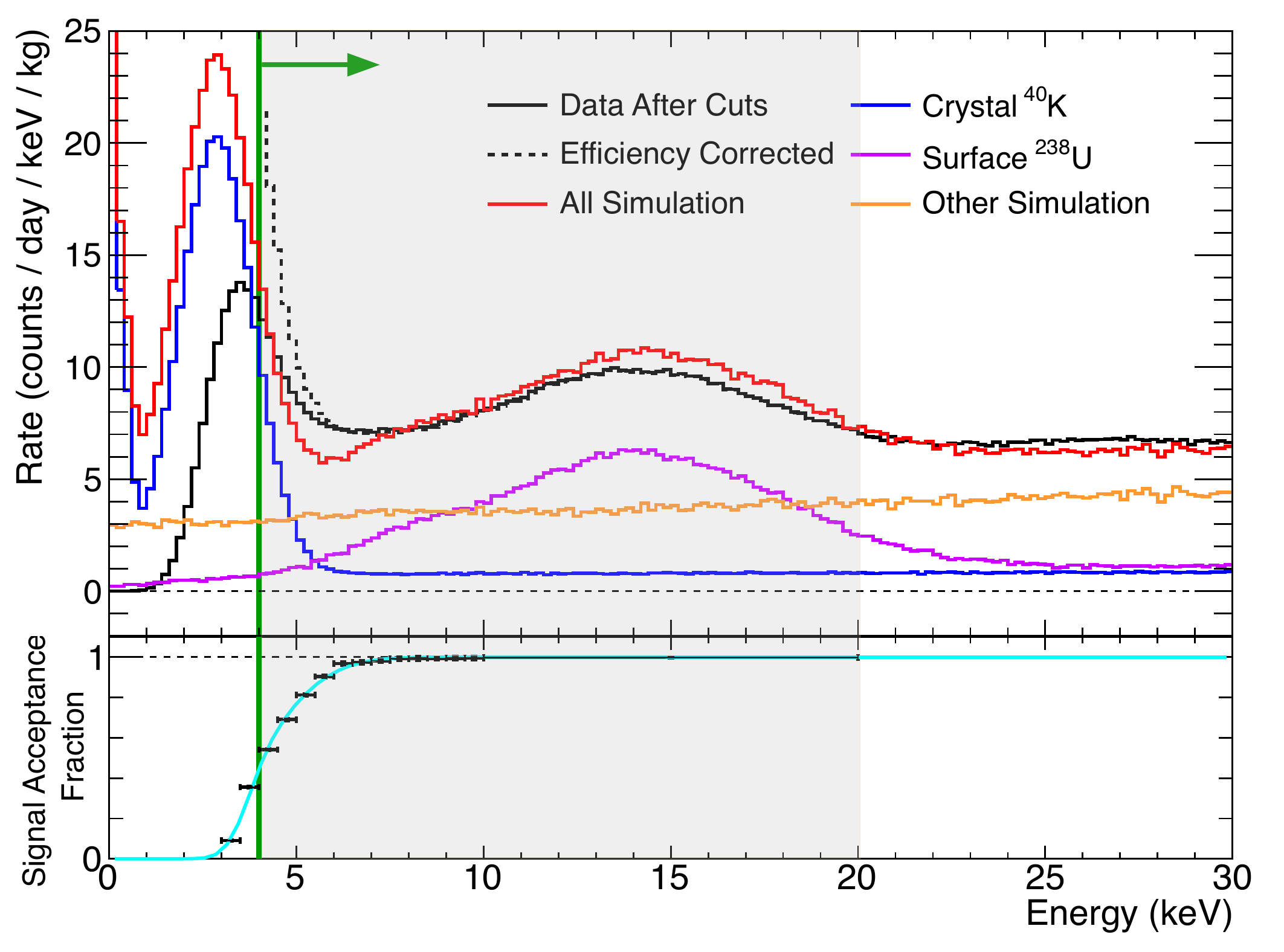}
		\caption{Energy spectrum of \dmi\ from 0\,--\,30\kev (top) and signal acceptance fraction (bottom). Shown is the spectrum from \dma\ after noise cuts (solid black) and with the addition of a cut efficiency correction (dashed black). The full simulation (red) is also shown, along with the contributions from \iso{K}{40} in the crystal (blue), \iso{U}{238} surface contamination on the copper encapsulation of the crystals (violet), and the sum total of backgrounds from other sources (orange). The signal acceptance fraction for the noise cuts is shown below the spectra, with the cut efficiency data (black dots) fit with a functional form (cyan) used in calculations. The green line indicates the 4\kev\ analysis threshold and the shaded area represents the region of interest (4--20 keV) in this analysis. \dmb\ data are similar, though its lower gain leads to less efficient noise cuts and higher analysis threshold (6\kev). }
		\label{lespec}
	\end{center}
\end{figure}

The spectrum in the region of interest (ROI) includes a 3\kev\ peak from \iso{K}{40} contamination in the crystals, a 15\kev\ feature from surface \iso{U}{238}-chain contamination, and a flat continuum from \mbox{0\,--\,20\kev} at 3\,--\,4\dru\ from other known contaminants in the detector components (see~\fig{lespec}). The 15\kev\ feature can be attributed mostly to \iso{U}{238}-chain contamination on the inner surface of the copper encapsulation, and this type of surface contamination has been observed previously in other NaI(Tl) experiments~\cite{ANAIS_tritium}. The simulation of \dmi~\cite{DMIce2014} predicts the general shape of these major features, though a number of additional cosmogenically activated components are expected to play a role. As described below, an analysis threshold of \analysisthresholdA\ for \dma\ and \analysisthresholdB\ for \dmb\ was implemented. The discrepancy between the simulated \iso{K}{40} and the measured peak is mainly due to the efficiency of signal retention during noise removal.

Events that are inconsistent with scintillation in NaI(Tl) are rejected in the analysis as noise. Three types of noise trigger the detector: electromagnetic interference (EMI) with the detector monitoring queries, coincident SPE pulses, and ``thin peaks'' that exhibit a decay time an order of magnitude quicker than scintillation~\cite{DMIce2014}. SPE events mainly come about from accidental coincidence of PMT dark noise and phosphorescence from crystal interactions~\cite{DMIceMuons}. Thin peaks are suspected to originate from Cherenkov light emission in the quartz light guides or PMT windows~\cite{Bernabei:2008yh, Amare:2014eea}, and frequently exhibit asymmetric pulse heights in the two PMTs. Thin peaks are the dominant noise source above the analysis threshold for this analysis.

Scintillation in NaI(Tl) has a \aprox250\nsec\ decay constant associated with photon emission, and photoelectrons arrive at discrete times over the entire readout. EMI events include no photoelectrons while SPE events include just one photoelectron per PMT. Only thin pulse noise events include multiple photoelectrons; however, all of these photoelectrons are observed as arriving at the same time. Therefore, events with photoelectrons arriving at multiple distinct times are distinguishable as scintillation. By requiring events to have four photopeaks in each PMT (peak finding cut), 70.8\plm0.3\% of scintillation events from 4\,--\,6\kev\ is accepted while producing a signal-to-noise ratio of 230\plm20. 

The peak finding cut efficiency is assessed via analysis of a secondary pulse shape discrimination parameter (pulse integral\,/\,pulse height). This parameter provides sufficiently independent discrimination capability between distinct populations that fitting the distribution before and after the peak finding cut enables determination of the peak finding cut efficiency in individual energy bins. Analysis threholds of 4\kev\ for \dma\ and 6\kev\ for \dmb\ were imposed. The signal acceptance fraction is shown in~\fig{lespec}. The large background rate from high concentration of \iso{K}{40} in these particular crystals largely limits the sensitivity of \dmi\ to dark matter at these energies. The behavior of this cut is consistent over time at all energies. No other cuts are employed in this analysis.

The ROI is split into four separate energy bins (\mbox{4\,--\,6}, 6\,--\,8, 8\,--\,10, and 10\,--\,20\kev), though the 6\kev\ analysis threshold of \dmb\ precludes analysis of the lowest energy bin in that detector. Therefore, seven total analysis energy bins exist between the two crystals, and the physics run is additionally split into 87 half-month time intervals. After correcting for live time and cut efficiency, the event rates in each energy bin decrease during the physics run by 8\%, 9\%, 8\%, and 6\% in the \dma\ event rates for the 4\,--\,6, 6\,--\,8, 8\,--\,10, and 10\,--\,20\kev\ bins respectively. For \dmb, the decrease is 8\%, 8\%, and 6\% for the 6\,--\,8, 8\,--\,10, and 10\,--\,20\kev\ bins. 

The total event rate in each energy bin linearly decreases as a function of time and it is accounted for in the fit,
\begin{equation}\label{fit}
	\textrm{Event Rate} = mx+b+A\cdot\cos{\frac{2\pi(x-x_{0})}{t_{0}}},
\end{equation}
where \textit{m} and \textit{b} are linear fit parameters, and \textit{A}, $x_{0}$, and $t_{0}$ are annual modulation fit parameters. The period of the fit, $t_{0}$, is fixed to one year (365.15 days) in this analysis.

\begin{figure}[t]
	\begin{center}
		\subfigure{
			\includegraphics[width=0.50\textwidth]{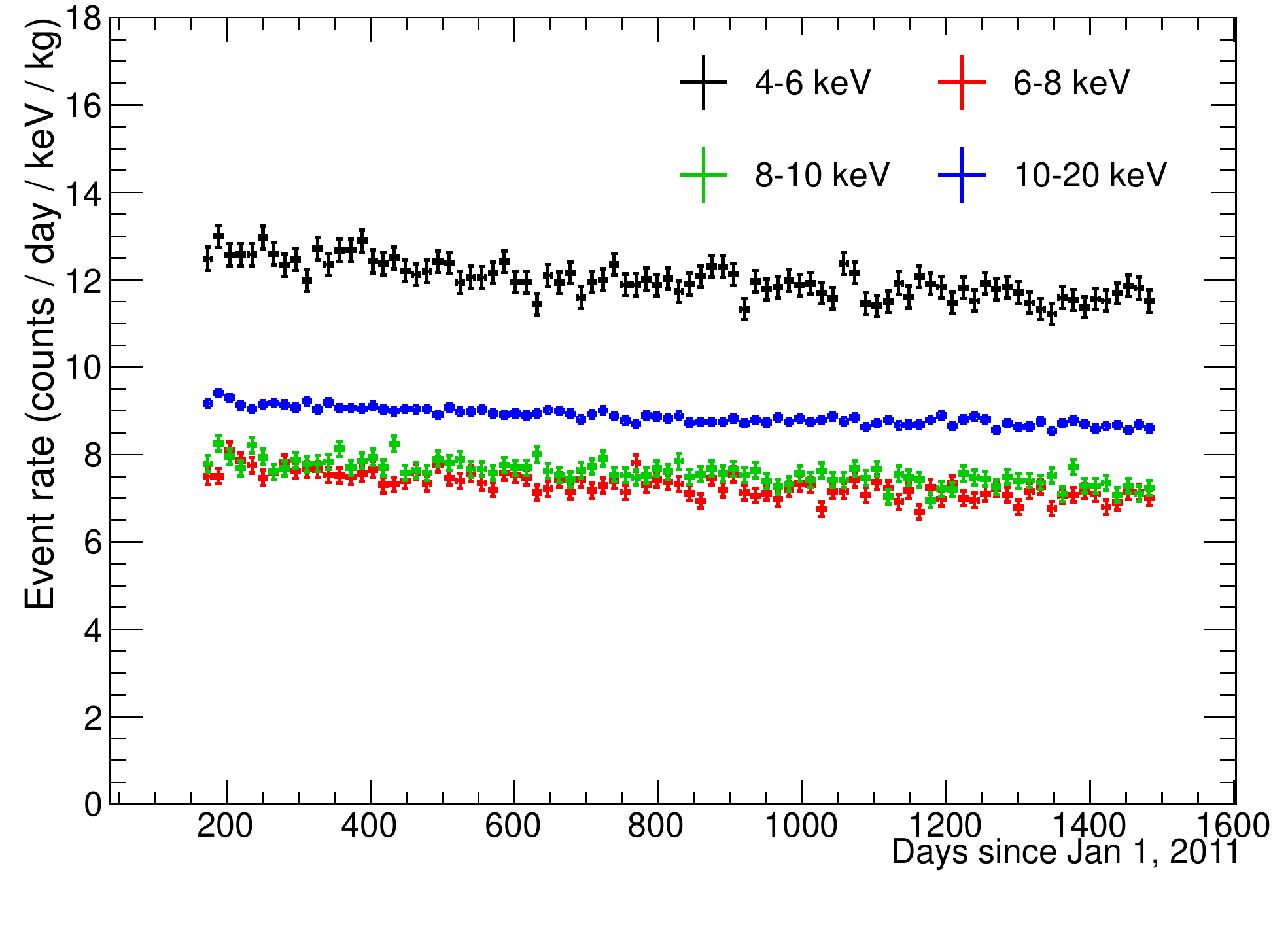}
		}
		\subfigure{
			\includegraphics[width=0.50\textwidth]{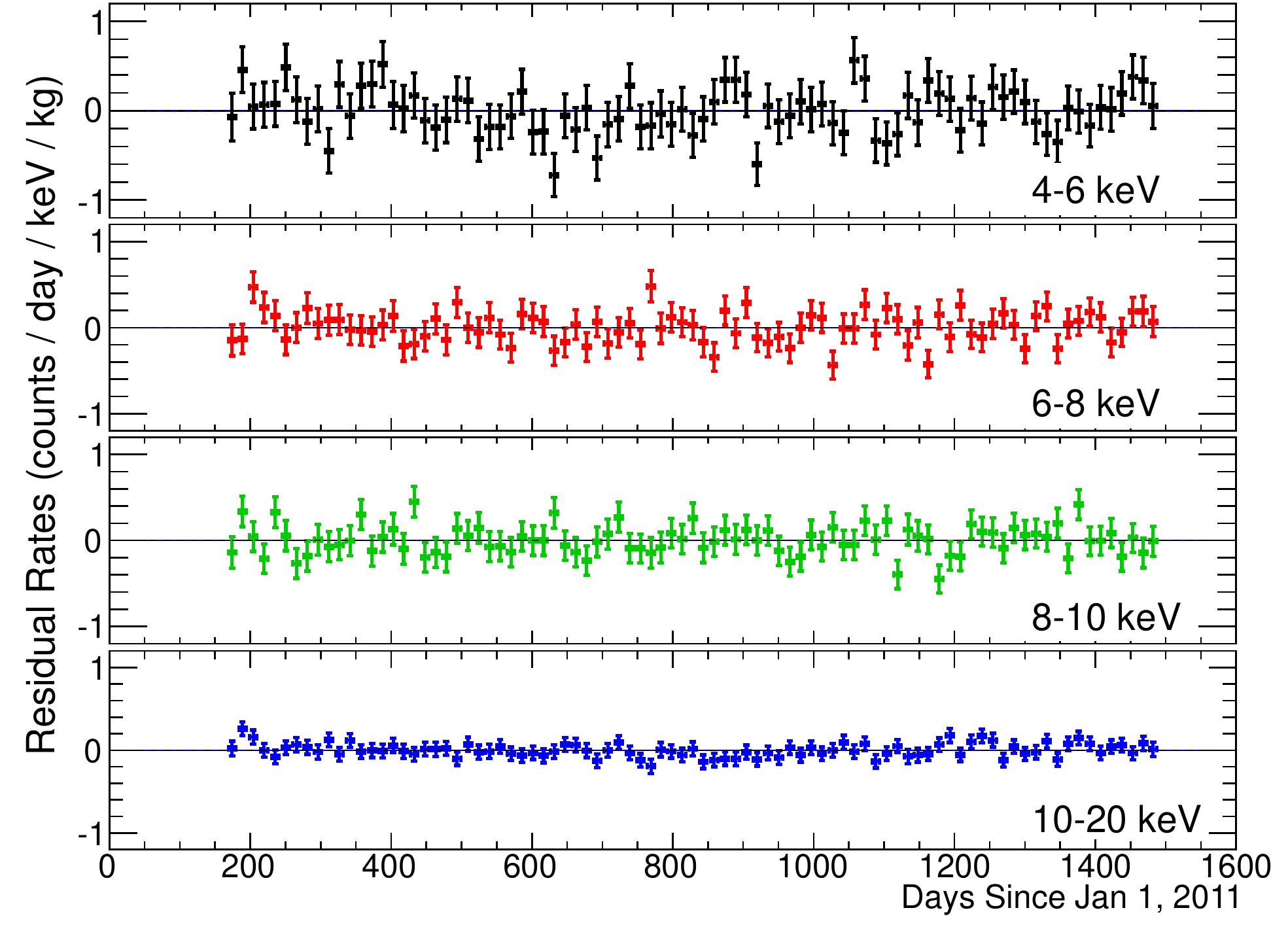}
		}
		\caption{Event rate vs.\ time for \dma\ before subtracting the linear rate component (top) and residual event rate vs.\ time after the subtraction (bottom). Rates are provided in half-month intervals for 4\,--\,6\kev\ (black), 6\,--\,8\kev\ (red), 8\,--\,10\kev\ (green), and 10\,--\,20 \kev\ (blue). The horizontal error bars represent the half-month bin width and the vertical error bars represent $\pm1\sigma$ error due to statistical and uptime uncertainties. In each of these energy bins, the data are consistent with the no modulation. \dmb\ data are similarly consistent with no modulation and exhibit a comparable linear rate component. }
		\label{evRates}
	\end{center}
\end{figure}

Cosmogenically activated isotopes (such as \iso{Co}{60} and \iso{I}{125}) and the broken \iso{Th}{232} and \iso{U}{238}-chains produce known changes in the event rate over time~\cite{DMIce2014, KIMS_NaI, Amare:2013lca}, though not enough to explain the decrease at the ROI energies. Cosmogenically activated \iso{H}{3} provides a possible source of the decrease~\cite{ANAIS_tritium}. The \iso{U}{238} surface contamination built into the \dmi\ background model~\cite{DMIce2014} is in equilibrium; however, a broken \iso{U}{238}-chain at \iso{Pb}{210} would produce similar spectral features while introducing decreasing event rates. The decay in the data rate is consistent with a combination of \iso{H}{3} and the broken \iso{U}{238}-chain, though a precise model requires further investigation.

\fig{evRates} shows the \dma\ event rates for each energy bin before (top) and after (bottom) subtracting off the fitted linear component. To perform the modulation analysis, a likelihood minimization fits the event rate over time for each energy bin with a sinusoid atop the linear background. By varying the number of free parameters, a variety of models can be tested. For the 4\,--\,6\kev\ bin of \dma, chi-squared analysis produces \ensuremath{\chi^2}/d.o.f. (p-value) = 86.11\,/\,87 (0.51) for the null hypothesis, \ensuremath{\chi^2}/d.o.f. (p-value) = 86.03\,/\,86 (0.48) for an annual modulation (fixed one year period) with the expected dark matter phase (fixed June 2nd maximum), and \ensuremath{\chi^2}/d.o.f. (p-value) = 84.35\,/\,85 (0.50) for an annual modulation (fixed one year period) with floating phase. The other energy bins are similarly consistent with the null hypothesis with p-values of 0.26 (6--8 keV), 0.60 (8--10 keV), and 0.55 (10--20 keV), providing no evidence for an annual modulation.

\begin{figure}[t]
	\begin{center}
		\includegraphics[width=0.50\textwidth]{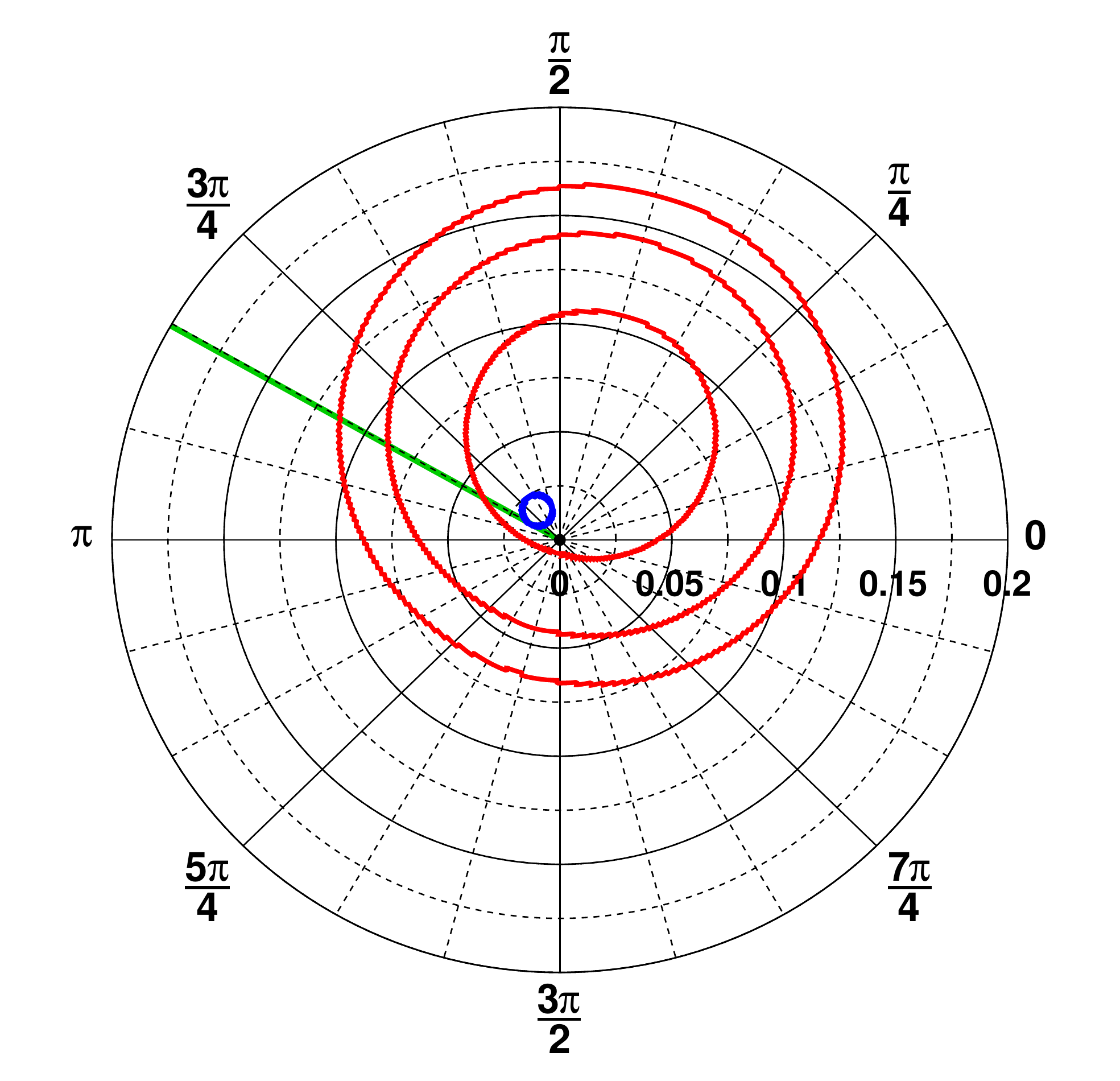}
		\caption{ Allowed regions in amplitude (\ensuremath{\mathrm{counts/day/keV/kg}}) vs. phase for annual modulation fits to the \dma\ 4\,--\,6\kev\ data, with contours at (inner to outer) 68\%, 95\%, and \mbox{99\% C.L (red)}. The DAMA/LIBRA 2\,--\,4\kev\ 99\%\,C.L.\ (blue) is also shown for comparison. Phase of 0 corresponds to January 1st. The predicted phase of a dark matter modulation signal under generic halo models June 2nd is indicated by the green line. }
		\label{polarLL}
	\end{center}
\end{figure}

The best fit to the \dma\ 4\,--\,6\kev\ bin has a modulation amplitude of 0.05\plm0.03\dru\ and a maximum on March 16th\plm42 days. DAMA has not published a floating phase best fit for 4\,--\,6\kev, so a direct comparison is not possible; however, across \mbox{2\,--\,6\kev}, DAMA/LIBRA observes a modulation amplitude of 0.011\plm0.001 and a best-fit phase of May 24th\plm7 days~\cite{DAMA_full}. A log-likelihood analysis comparing annual modulations of each amplitude and phase to the best fit shows that the data from \dmi\ are consistent with the null hypothesis (see~\fig{polarLL}). The limitations of this detector are also apparent as the DAMA/LIBRA 99\%\,C.L.\ contour is indistinguishable from the null hypothesis at the 68\% C.L.

\begin{figure}[!hb]
	\begin{center}
		\includegraphics[width=0.50\textwidth]{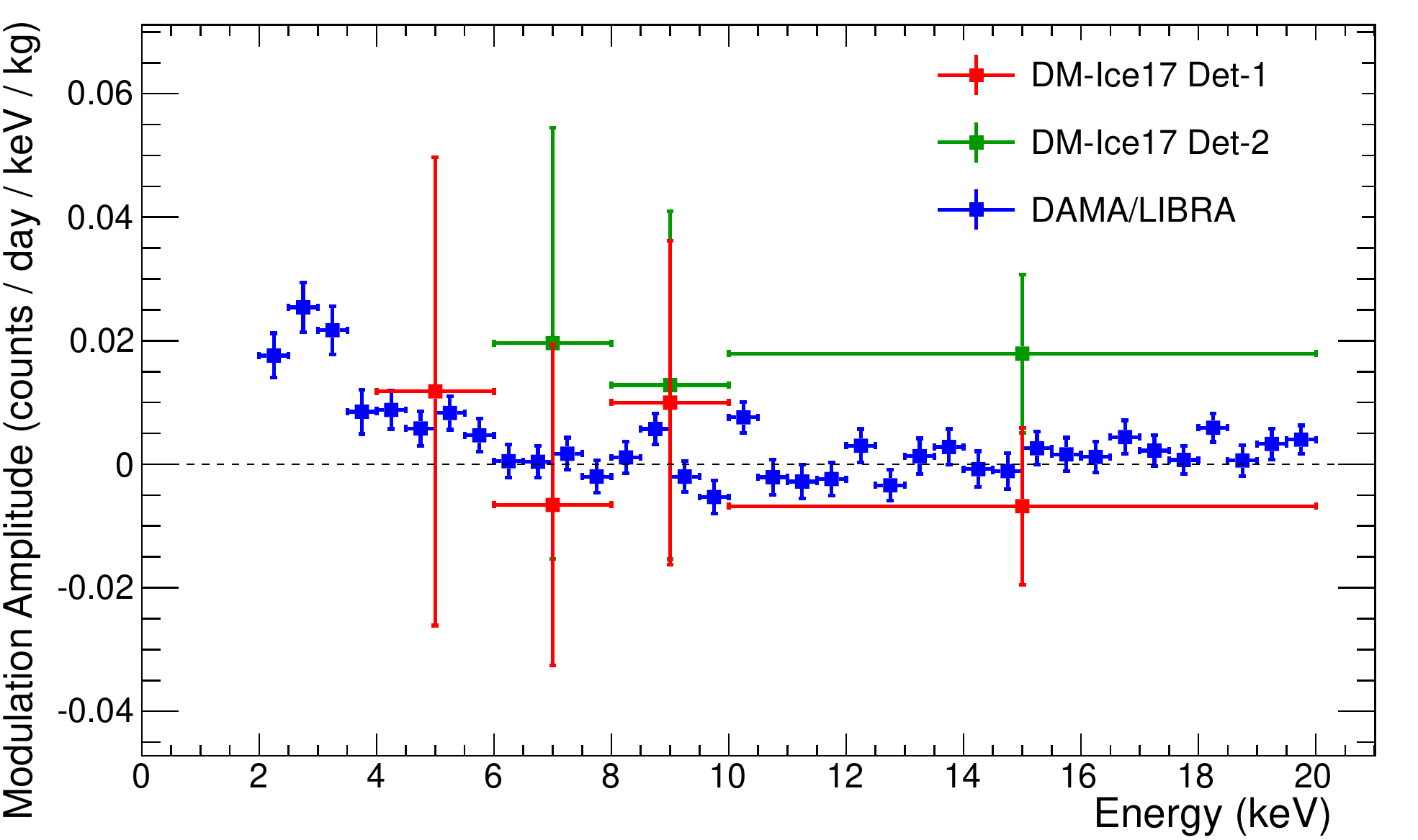}
		\caption{Amplitude of modulation vs energy showing maximum likelihood fits for DAMA~\cite{DAMA_full} (blue) and \dmi\ rates [\dma\ (red) and \dmb\ (green)]. The linear backgrounds underlying the event rate and the modulation amplitude are free parameters in these fits, with period and phase forced to that of an expected dark matter signal (1 year and 152.5 days, respectively). Horizontal error bars represent the width of the energy bins used for the analysis. Vertical error bars are $\pm 1\sigma$ error on the binned modulation fit amplitudes. }
		\label{modamp}
	\end{center}
\end{figure}

\begin{figure}[!t]
	\begin{center}
		\includegraphics[width=0.50\textwidth]{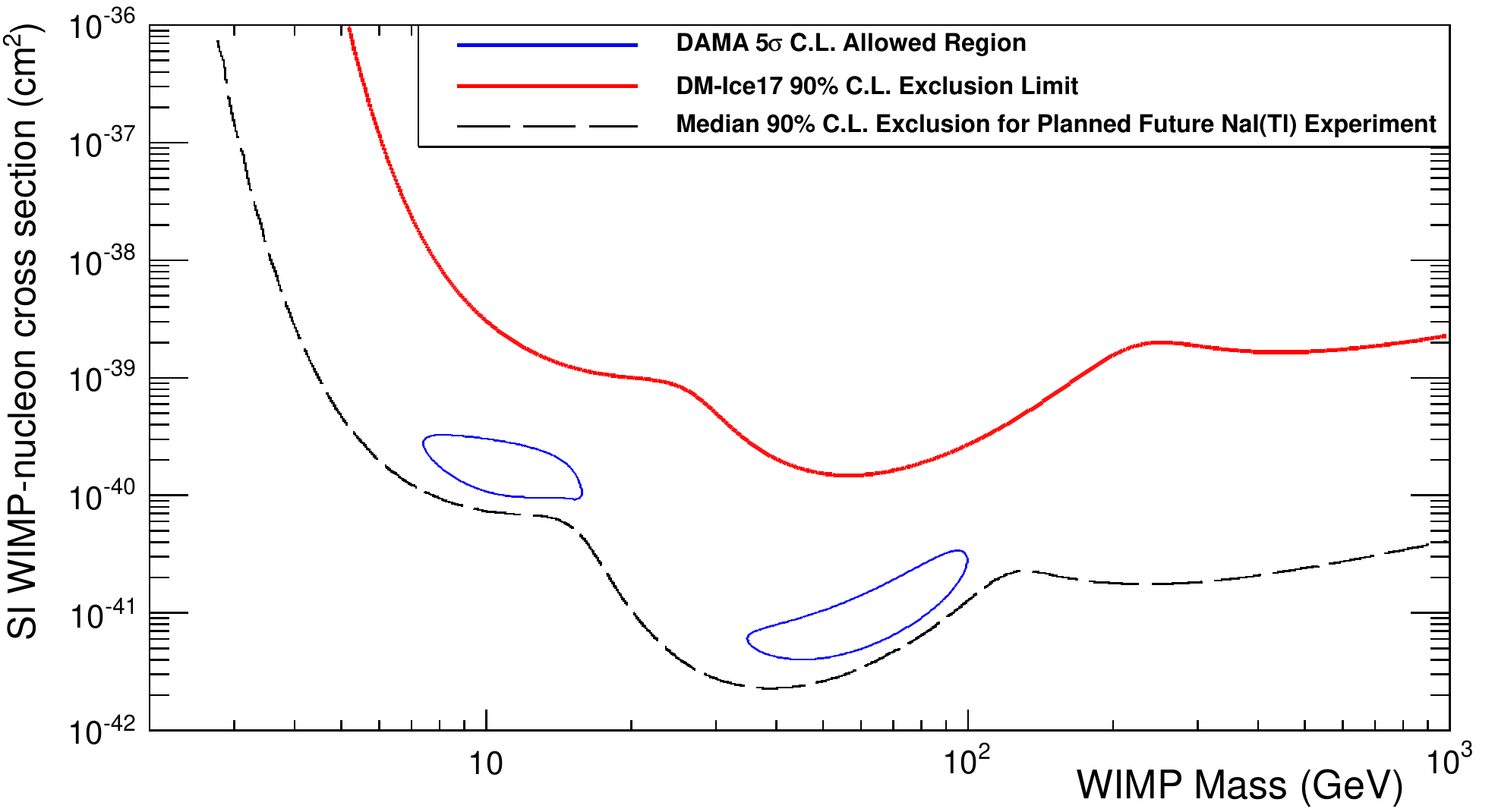}
		\caption{WIMP exclusion limits at 90\%\,C.L.\ from the \exposurekgyrs\ \dmi\ physics data set (red), with DAMA preferred $5\sigma$\,C.L.\ contour (blue) for comparison. As a reference, the projected sensitivity (median 90\%\,C.L.) for a planned future NaI(Tl)-based detector with 2\dru\ backgrounds in the ROI after 500\kgyr\ and a 2\kev\ analysis threshold is shown in the dashed black line~\cite{Zack}.}
		\label{exclusion}
	\end{center}
\end{figure}

Modulation fits with fixed period of one year and fixed phase of 152.5 days are consistent with zero amplitude (see~\fig{modamp}). The best-fit modulation amplitudes across the entire ROI can be combined to set limits in the WIMP parameter space assuming a standard halo model with WIMP density of 0.3\gev, disk rotation speed of 220\kms, Earth orbital speed of 29.8\kms, and galactic escape velocity of 650\kms. An exclusion limit (see~\fig{exclusion}) is produced via a log-likelihood analysis of the observed binned modulation amplitudes, with predicted modulation amplitudes for particular WIMP candidates as described in ~\cite{Freese}. Best-fit contours for the full DAMA/LIBRA-phase1 run~\cite{DAMA_full} were produced along with the \dmi\ exclusion limit for comparison, based on the methodology from~\cite{Savage_1}. Uncertainties in WIMP-nucleon coupling to different targets do not play a role in the comparison, since \dmi\ and DAMA share the NaI(Tl) target material.

The exclusion limits and best-fit contours shown here are produced via a log-likelihood analysis rather than via the goodness-of-fit method, both methods described in~\cite{Savage_1}. We find that while generally similar, the goodness-of-fit methodology introduces an undesirable dependence on the energy binning not present in the log-likelihood method. We perform a log-likelihood analysis on each data set to avoid introducing effects due to differences in energy binning between the DAMA and \dmi\ data.

This analysis constitutes the strongest limit set in the Southern Hemisphere by a direct detection dark matter search, an important step towards an unambiguous statement about the nature of the DAMA signal. \dmi\ is limited by its high internal backgrounds and low mass; environmental backgrounds, detector stability, and difficulties inherent to remote deployment do not limit the capabilities of \dmi. Recent research efforts have yielded crystals with better radiopurity~\cite{Pettus, Amare:2013lca, KIMSCrystalRD}. A planned future NaI(Tl) experiment with 2\dru\ backgrounds in the ROI, a 500\kgyr\ exposure, and a 2\kev\ analysis threshold can definitively test DAMA, as shown in~\fig{exclusion}.

In summary, we report the results from the first search for an annual modulation dark matter signal in the Southern Hemisphere using NaI(Tl) as the target material. \dmi\ establishes the South Pole as a site for underground low-background experiments. The \dmi\ data, taken over 3.6 years for a total exposure of \exposurekgyrs, show no evidence of annual modulation in the 4\,--\,20\kev\ energy range. This yields the strongest limit from a direct detection Southern Hemisphere dark matter search, establishing a solid foundation for future searches.

\begin{acknowledgments}
We thank the Wisconsin IceCube Particle Astrophysics Center (WIPAC) and the IceCube Collaboration for their ongoing experimental support and data management. We thank Chris Toth and Emma Meehan for operational support at Boulby Underground Lab. This work was supported in part by the Alfred P.\ Sloan Foundation Fellowship, NSF Grants No.~PLR-1046816, No.~PHY-1151795, and No.~PHY-1457995, WIPAC, the Wisconsin Alumni Research Foundation, Yale University, the Natural Sciences and Engineering Research Council of Canada, and Fermilab, operated by Fermi Research Alliance, LLC under Contract No.~DE-AC02-07CH11359 with the United States Department of Energy. W.~P.~and A.~H.~were supported by the DOE/NNSA Stewardship Science Graduate Fellowship (Grant No.~DE-FC52-08NA28752) and NSF Graduate Research Fellowship (Grant No.~DGE-1256259) respectively. 

\end{acknowledgments}%


\addcontentsline{toc}{section}{References}

\bibliographystyle{apsrev4-1}

\bibliography{%
intro,%
dmice,%
else%
}

\end{document}